# High Order M-QAM Massive MIMO Detector with Low Computational Complexity for 5G Systems

Vishnupraneeth P, Pravir Singh Gupta, and Gwan Seong Choi.

*Abstract*—In this work, the behaviour of bit error rates for both conventional and massive MIMO systems with high order constellations, which are essential to achieve spectral efficiency for 5G standard communications, has been evaluated. We have used real-domain Schnorr Euchner enumeration with $K$-best algorithm to reduce computational complexity of detection. The results, presented in this letter, have outperformed existing detection algorithms in terms of complexity and BER, especially in low SNR regions, for both massive and conventional MIMO systems. We performed simulations for $N \times N$ MIMO system, where $N = 8, 25, 40, 50, 60, 80, 100$ and $120$, for both $256$-QAM and $1024$-QAM high order transmission systems as per the latest 3GPP standards for 5G systems and beyond. All the analyses and results that are given in this letter are from our MIMO detector, prior to usage of error correction decoder.

*Index Terms*—1024-QAM, 256-QAM, Massive MIMO systems, Low SNR-channel, Low computational complex algorithms, Schnorr Euchner enumeration, K-best algorithm.

## I. INTRODUCTION

Massive number of spatial multiplexing data streams integrated with high order constellations is the key to building efficient future wireless technology which will meet the bandwidth and transmission speed requirements of modern day communications [1]. Low order constellations consume more amount of energy to transmit same amount of information, compared to high orders, which transmit more bits per each constellation symbol. Multi Input Multi Output (MIMO) systems have become the necessary standard for building efficient wireless technologies such as Long Term Evolution (LTE), IEEE 802.11/802.16 [2] and is also requirement for $5G$ wireless systems. Energy consumed at the receiver end, to retrieve the input signals sent from base station (BS) has introduced major limitations for using massive MIMO systems or high order quadrature amplitude modulation ($M$-QAM) transmission standards. Hence low computationally complex algorithm achieving performance similar to a near Maximum Likelihood (Near-ML) detector for massive MIMO is required to transmit in the range of higher spectral frequencies and required bandwidth. Many existing detection algorithms like Minimum Mean Square Error (MMSE), Zero Forcing (ZF), Successive interference cancellation (SIC) [3] may be used for lower dimensional MIMO systems and lower order constellations. The conventional breadth first, depth first search, list first search or hybrid MMSE K-best algorithms [4] also displays extensive computational complexity when dealing with higher dimensional systems or higher order QAM transmission standards. Linear approximation detection algorithm as proposed in [5], splits N number of antennas into N/2 sub-systems to locally reduce the complexity for massive MIMO systems. There is a need for powerful detection algorithm to uncover the symbols, constructed from high order constellations, whilst retaining the performance, achieved by lower order constellations, especially for spatial multiplexing input streams. $K$-best algorithm integrated with Schnorr Euchner enumeration was proposed before, in [2], [6] but [2], [6], [7] evaluated this algorithm, only for low order $M$-QAM and conventional MIMO systems. Different algorithms [5], [8]–[10] were used for detection of higher order constellations but suffered from high computational complexity.

Since complexity of detection for massive MIMO systems has always been a concern, in this letter, we have studied the performance of Schnorr Euchner enumeration based detection algorithm [2], which provides extremely low computational complexity. We have simulated this algorithm for massive MIMO systems with different antenna configurations and for high order transmissions, as per latest 3GPP standards for 5G systems and beyond [11], which is also described in Table I. We have achieved better bit error rates, especially in low signal to noise ratio (SNR) regions, and similar performance in high SNR regions, thus fortifying the compatibility of this extremely low complex algorithm, for massive MIMO systems. The complexity analysis of algorithm used in this work, has also been discussed in detail, comparing it to various existing algorithms like conventional K-best, ML etc.

## II. SYSTEM MODEL

Let us assume a $(N_T \times N_R)$ MIMO system, transmitting with $M$-QAM constellation symbols. The basic MIMO system [8] can be modelled as

$$y = Hx + n \qquad (1)$$

where $y$ is $(N_R \times 1)$ received complex vector, $H$ is path gain $(N_R \times N_T)$ channel matrix, $x$ is $(N_T \times 1)$ input data

The Manuscript for this letter was submitted on ¡date¿. All the authors are from Electrical and Computer Engineering Department, Texas A&M University, College station, TX, USA.

Vishnupraneeth.P is currently pursuing MS in Computer Engineering at Texas A&M University, under supervision of Dr Gwan Choi. His area of research is VLSI for wireless communications and he is currently exploring various low complexity designs for Massive MIMO Detection.(e-mail: $vish.nu\_123@tamu.edu$).

Pravir Singh Gupta is currently pursuing PhD in Computer Engineering at Texas A&M University, under supervision of Dr. Gwan Choi. (e-mail: $pravir@tamu.edu$).

Dr Gwan Seong Choi is member of faculty at Department of Electrical and Computer Engineering, Texas A&M University, since 1994. (e-mail: $gwanchoi@tamu.edu$).

TABLE I
SYSTEM MODEL CONFIGURATION AND ANTENNA DIMENSIONS

| System Models | Antenna Configurations Used ($N_T \times N_R$) | Transmission Order of $M$-QAM | $K$ |
|---|---|---|---|
| Experiment Model 1 | $(8 \times 8), (25 \times 25),$ $(40 \times 40), (50 \times 50),$ $(60 \times 60), (80 \times 80),$ $(100 \times 100)$ and $(120 \times 120)$ MIMO systems | 256-QAM, 1024-QAM | 5 |
| Experiment Model 2 | $(8 \times 8)$ Conventional MIMO System, for different values of $K$-paths at each antenna level | 256-QAM, 1024-QAM | 5,10, 15,20, 100. |
| Experiment Model 3 | $(100 \times 100)$ Massive MIMO System, for different values of $K$-paths at each antenna level | 256-QAM, 1024-QAM | 5,10, 15,20. |

Where $M$ is the constellation order, $N_T$ represents number of transmit antennas, $K$ represents possible number of $K$-best paths at each level symbol matrix and $n$ as $(N_R \times 1)$ complex white Gaussian noise vector, which is derived from SNR.

Massive MIMO systems, as configured in Table I, decreases overall communication latency, in comparison to using high number of frames with conventional MIMO systems, for transmitting same amount of data. Further, using high order constellations increases the overall throughput and spectral efficiency of the system, by transmitting same amount of data with reduced energy consumption. For massive MIMO systems transmitting with high order constellation, extensive computations are required for powerful and accurate detection of symbols, which accounts for increase in energy consumption at the receiver end. However, with detection algorithm used in this letter, a quantitative trade-off can be achieved between energy consumption and spectral efficiency for massive MIMO systems with high order transmission, depending on requirement of BER, latency for various communications.

## III. SCHNORR EUCHNER BASED K-BEST ALGORITHM FOR 1024-QAM, 256-QAM

Conventional $K$-best algorithm searches all the possible symbols on each level, for all receive antennas to detect the symbol vector with lowest partial euclidean distance (PED). This complexity increases rapidly for higher order of constellations or massive MIMO systems. The number of symbols that are expanded to calculate PED can be measured as basis for computational complexity, to compare various algorithms. $K$-best algorithm, used with $M$-QAM transmission system expands $K \times \sqrt{M} \times 2N_T$ symbols in real domain and $K \times M \times N_T$ in complex domain for detecting complete input vector. As proposed in [2], using Schnorr euchner with $K$-best reduces this complexity to $(2K-1) \times 2N_T$ in real domain and $(2K-1) \times N_T$ in complex domain. This greatly takes effect for massive MIMO systems or higher order constellations. This can be clearly observed in Table II and Table III, where complexity of various algorithms are compared, for Experiment Model 2 and Experiment Model 3. The general complexity analysis for massive MIMO systems has been studied in Table IV. Any $N_T \times N_R$ complex system can be converted to $2N_T \times 2N_R$ real system and hence the number of levels to detect using $K$-best will also double for real domain. In general, for detection at any level of the tree, each $K$-best parent from previous level will be expanded to their $\sqrt{M}$ possibilities in real domain and $M$ possibilities in complex domain [6]. Then, $K$-best nodes for current level are selected using PED as metric. Using Schnorr Euchner row enumeration, the symbol with lowest PED, termed as first child and the adjacent symbols with next lowest PED, termed as next child are expanded for each $K$-best parent from previous level [6]. The $H$ matrix $(N_R \times N_T)$ is transformed into upper triangular matrix $R$ $(N_T \times N_T)$ by QR decomposition (CORDIC algorithm using Givens rotations [12]) and the transpose of resulting $Q$ matrix $(N_R \times N_T)$ is multiplied with the system model equation defined in eq. (1), transforming it equation to

$$Y = Rx + N \qquad (2)$$

where $Y$ ($= Q' \times y$) is $(N_T \times 1)$ complex received vector, $R$ is $(N_T \times N_T)$ upper triangular matrix, $x$ is $(N_T \times 1)$ input data symbol matrix and $N$ ($= Q' \times n$) is $(N_T \times 1)$ complex white Gaussian noise vector.

Starting from the last level of $Y$ matrix in eq. (2), the first child is measured, assuming zero noise. This symbol is added to the $K$-list for this current level. Another $K-1$ next child symbols are measured by Schnorr euchner row enumeration, which picks the adjacent symbols to the first child alternatively on either side, in the real domain. Symbols in $K$-list of last level act as parents for the penultimate level. The first child for each of the symbol is calculated assuming zero noise, and with their PEDs, all $K$ first childs are sorted, which is termed as distributed sorting, contrary to global sorting [6], implemented in conventional algorithm. The first child with lowest PED is added to $K$-list for penultimate level, and the next child of corresponding parent is added to current sorter list, for this level, replacing the first child. This list is again sorted, and the best child with lowest PED is added second to $K$-list for penultimate level, and the next child of corresponding parent is added to current sorter list, replacing best child, that was added to $K$-list for penultimate level. This algorithm is iterated till all $K$ symbols are loaded into the $K$-list for penultimate level. And this process is iterated for each receiver antenna, starting from bottom row. For massive MIMO $(100 \times 100)$ systems with high order constellations, all the symbols, corresponding to 100 receiver antennas are detected, each at a time, using Schnorr Euchner row enumeration, starting from bottom level. For massive MIMO system, the complexity for 256-QAM transmission, depends on the value of $K$ that is usually set high for higher order constellations. Table II and Table III, shows the complexity of different algorithms, for different values of $K$, in terms of number of nodes processed, since it is the computationally extensive part of any detector and can affect latency of total system.

From the complexity analysis in Table IV, it can be clearly deduced that as number of antennas increase the complexity of MIMO system increases exponentially for ML algorithm, increases linearly for $K$-best algorithms. It can also be noticed that complexity of the Schnorr Euchner based $K$-best is inde-

TABLE II
COMPLEXITY ANALYSIS FOR MASSIVE MIMO SYSTEMS WITH HIGH ORDER M-QAM

| Massive MIMO ($100 \times 100$) Detection Algorithm | 256-QAM (K=5) | 1024-QAM (K=5) | 256-QAM (K=10) | 1024-QAM (K=10) |
|---|---|---|---|---|
| ML | $256^{100}$ | $256^{125}$ | $256^{100}$ | $256^{125}$ |
| Conventional $K$-best [6] | $10^{5.5}$ | $10^{6.0}$ | $10^{5.7}$ | $10^{6.3}$ |
| This work | $10^{3.2}$ | $10^{3.2}$ | $10^{3.5}$ | $10^{3.5}$ |

Each value represents number of nodes expanded for each corresponding algorithm, during detection.

TABLE III
COMPLEXITY ANALYSIS FOR CONVENTIONAL MIMO SYSTEMS WITH HIGH ORDER M-QAM

| Conventional MIMO ($8 \times 8$) Detection Algorithm | 256-QAM (K=5) | 1024-QAM (K=5) | 256-QAM (K=10) | 1024-QAM (K=10) |
|---|---|---|---|---|
| ML | $256^{8}$ | $256^{10}$ | $256^{8}$ | $256^{10}$ |
| Conventional $K$-best [6] | $10^{4.3}$ | $10^{4.9}$ | $10^{4.6}$ | $10^{5.2}$ |
| This work | $10^{2.1}$ | $10^{2.1}$ | $10^{2.5}$ | $10^{2.5}$ |

Each value represents number of nodes expanded for each corresponding algorithm, during detection.

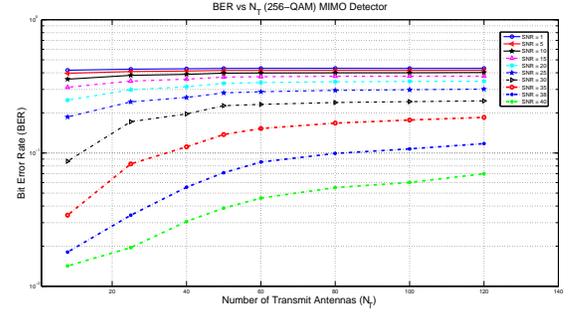

Fig. 1. Performance evaluation of $BER$ by increasing number of transmit antennas for different $SNR$ regions (256-QAM Transmission).

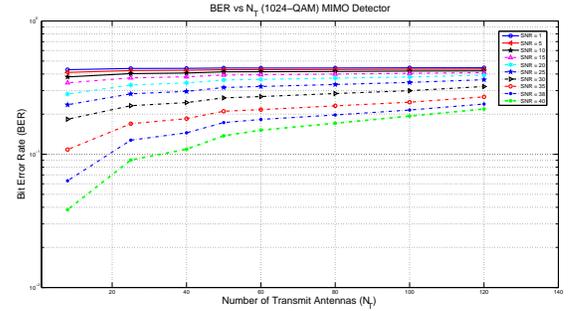

Fig. 2. Performance evaluation of $BER$ by increasing number of transmit antennas for different $SNR$ regions (1024-QAM Transmission).

pendent of constellation order. Even though, higher value of $K$ may be required to achieve better BER results for higher order constellations, the increase in complexity is linear, for both massive and conventional MIMO systems. Our results display better BER performance than recent papers [5], [9], [10], especially for low SNR channels and similar performance for high SNR channels, with much less computational complexity. Log likelihood ratio (LLR) values are measured for above models, which are directed into our Error Correcting (ECC) decoder, for our following work, which should further improve our BER performance.

## IV. SIMULATION RESULTS AND ANALYSIS

System models described in Table I, were used to produce evaluation results. We have simulated ten thousand frames of input bit streams, with each frame transmitting $N_T$ symbols, which were generated randomly. For our Experiment Model 1, as described in Table I, we have studied the dependency of BER on conventional and massive MIMO system dimensions, transmitted with high order constellations. We have utilized

TABLE IV
GENERAL COMPLEXITY ANALYSIS

| ML | Conventional $K$-best [6] | This work |
|---|---|---|
| $M^{N_T}$ | $K \times \sqrt{M} \times 2N_T$ | $(K + K - 1) \times 2N_T$ |

Where $M$ is the constellation order, $N_T$ represents number of transmit antennas, $K$ represents possible number of $K$-best paths at each level. This table compares the general complexity of algorithm used [2] in this letter, with other detection algorithms. Each value represents number of nodes expanded for each corresponding algorithm, during detection

Experiment Model 2 and Experiment Model 3, as described in Table I, to analyse the behaviour of BER for different range of low and high SNR regions, using different values of possible $K$-best paths at each antenna level. We have selected one antenna configuration to represent each of conventional and massive MIMO systems and study behaviour of BER with respect to SNR. We juxtaposed high order constellations of above systems, using different values of $K$ discretely, to analyse dependency of BER on SNR, and also on the antenna dimensions of massive MIMO system. The BER results displayed in this letter, are derived just from the MIMO detector, prior to using out ECC iterative decoder, which will be presented in our next work.

The graphs shown in Fig. 1 and Fig. 2 presents the performance of Schnorr Euchner based $K$-best algorithm, for different number of transmit antennas, using 256-QAM and 1024-QAM, for various SNR channels. The graph is on semilog scale, with BER values presented on logarithmic scale.

It can be deduced from Fig. 1 and Fig. 2 that at low SNR i.e. around SNR $< 10$, the performance of this low complexity detector is almost identical for different dimensions of MIMO system. However for high SNR i.e. SNR $> 10$, BER increases with increase in dimension of MIMO system. This can be the due to the fact that bits in input bit stream, sent from base station are more vulnerable to get flipped in high SNR regions, when we try to increase the dimensions. For low SNR regions, the bit stream is equally deteriorated due to very high noise power resulting in identical performance of the detector with the number of transmit antennas.

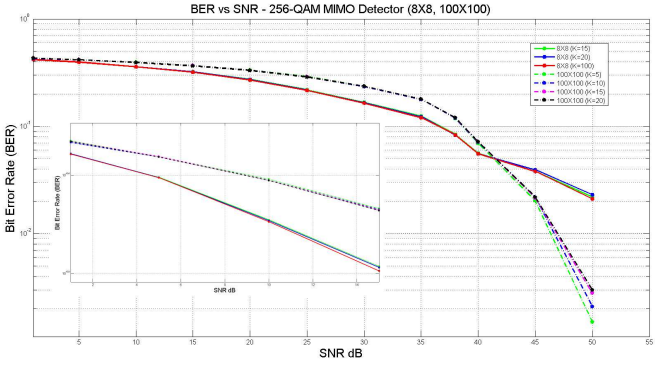

Fig. 3. Behaviour of $BER$ with respect to $SNR$ for Massive and Conventional $MIMO$ systems, with 256-QAM, for different $K$-best paths.

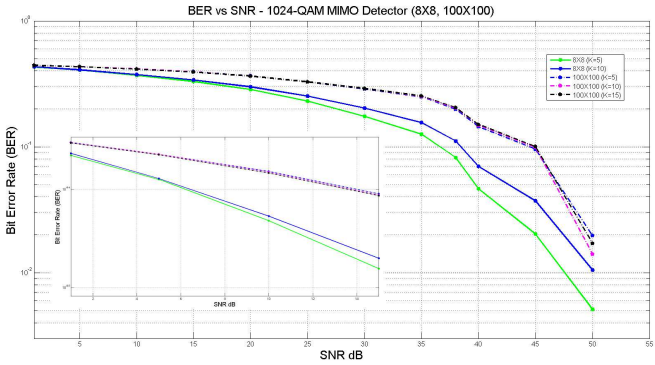

Fig. 4. Behaviour of $BER$ with respect to $SNR$ for Massive and Conventional $MIMO$ systems, with 1024-QAM, for different $K$-best paths.

The BER vs. SNR graphs shown in Fig. 3 and Fig. 4 represent $(8 \times 8)$ and $(100 \times 100)$ systems with 256 and 1024-QAM transmitting configuration respectively. The configuration of these experiments appear in Table I as Experiment Model 2 and Experiment Model 3.

It can be observed from Fig. 3 and Fig. 4 that increasing number of $K$-best paths per receiver antenna, decreases BER, as expected, especially in low SNR regions. This is due to high noise power, compared to signal power and possible high scale of deviation for input constellation symbols, resulting in increased difficulty to detect received vectors. Increasing $K$-best paths increases the number of candidate solutions which in turn increases the probability of detecting the transmitted input symbols accurately. However, this behaviour of BER is unanticipated for high SNR regions and this improvement is not noticeable, with increase in possible $K$-best paths. This is due to the fact that input symbols are deviated less, in high SNR regions and hence the transmitted symbol can be accurately guessed from less number of candidate solutions implying acceptable lesser $K$ value for $K$-best paths at each antenna level.

## V. CONCLUSION

For massive MIMO systems transmitting with high order constellation, we have evaluated performance and analysed the complexity of extremely low computationally complex, real domain Schnorr Euchner enumeration detector. We have generated results for $N \times N$ MIMO system, where $N = 8, 25, 40, 50, 60, 80, 100$ and $120$, for both 256-QAM and 1024-QAM. For very low SNR regions, bit error rate remains non-fluctuating as transmit antennas are increased, but BER increases at decreasing rate, for high SNR regions. Increasing number of $K$ paths, moderately decreases BER, at low SNR regions due to high possibility of bit flip in channel and high possibility of improvement as $K$ radius at each level is increased. However, for high SNR regions, low $K$ radius is adequate, for detecting majority of bits that can be corrected, without using decoding. In future work, we will use ECC decoder, along with low complexity detector, used in this letter, to further decrease the BER of the massive MIMO systems, with high order transmission systems, especially in low SNR regions.